\documentclass[journal]{IEEEtran}

\usepackage{cite}
\usepackage{amsmath,amssymb,amsfonts}
\usepackage{algorithmic}
\usepackage{graphicx}
\usepackage{textcomp}
\usepackage{xcolor}
\usepackage{booktabs} 
\usepackage{url}
\usepackage{hyperref}
\usepackage{multirow} 
\usepackage{array}
\usepackage{makecell}
\usepackage{tabularx}

\newcommand{\Rpu}{R_{pu}}
\newcommand{\Tg}{\ensuremath{T_{g}}}

\begin{document}
	
	\title{Modal Analysis of Core Inertial Dynamics: Re-evaluating Grid-Forming Control Design Principles}
	
	\author{Gerardo~Medrano and Santiago~C\'obreces
		\thanks{The authors are with the Department of Electrical Engineering.}
	}
	
	\maketitle
	
	\begin{abstract}
		This paper employs modal analysis to study the core inertial dynamics of governor-controlled synchronous generators (GC-SG), droop-based grid-forming (GFM) converters, and their most fundamental interactions. The results indicate that even in the simplest cases, the prevailing industry paradigm of emulating legacy GC-SG behaviour in GFM converters (high inertia to slow down the system and large droop to increase damping) could be a suboptimal policy. It is shown that GC-SGs exhibit a fundamental trade-off: adequate damping of the turbine-governor mode requires large droop constants, inevitably increasing steady-state frequency deviation and dependence on secondary regulation. In contrast, droop-based GFM converters invert this relationship: decreasing the droop constant simultaneously reduces steady-state frequency deviations and increases damping, while allowing virtual inertia to be freely chosen. When two GC-SGs are coupled, the poorly damped electromechanical swing mode emerges. Results show that replacing one GC-SG with a GFM converter of equivalent droop and inertia already significantly improves damping of both swing and turbine-governor modes. Counter-intuitively, further and remarkable damping gains are achieved by substantially lowering the GFM virtual inertia constant. These findings suggest that current industry trends may be constraining the potential benefits of Inverter Based Resources (IBRs). Optimal stability and performance are instead obtained with low droop and low virtual inertia, yielding tightly bounded frequency variations and strongly-damped electromechanical modes. The results indicate a need to re-evaluate GFM control design principles and emerging grid-code requirements.
	\end{abstract}
	
	
	\section{Introduction}
	\IEEEPARstart{T}{he} global energy transition toward decarbonization has dramatically increased the penetration of renewable energy sources interfaced through power electronic converters to conventional and isolated power systems \cite{Blaabjerg2023,Milano2018}. This shift progressively displaces conventional synchronous generators with inverter-based resources (IBR), significantly reducing total system inertia and damping while introducing new small-signal stability challenges related to frequency nadir, rate-of-change-of-frequency (RoCoF), and electromechanical oscillations \cite{Milano2018,Markovic2021,Lin2020}.
	
	Grid-forming (GFM) converters have emerged as a critical technology to enable stable operation of such systems \cite{Pogaku2007,Zhong2016}. Unlike grid-following inverters that require a stiff voltage reference, GFM units can independently form voltage and frequency, facilitating black-start capability and autonomous, islanded operation. Most GFM implementations employ droop-based control or virtual synchronous machine (VSM) strategies that intentionally emulate the response of synchronous generators, thereby providing ``virtual inertia'' and ``virtual damping'' to the network \cite{Zhong2016,Rocabert2012,Poolla2022,Zhong2011}. This emulation paradigm has dominated research, demonstration projects, and emerging grid codes, which often mandate that IBR must deliver a minimum virtual inertia constant and follow synchronous-machine-like response \cite{Lin2020,UNIFI2024,VDE2025,AEMO2023,ENTSOE2024}.
	
	In conventional GC-SGs, sufficient damping of the low-frequency turbine-governor mode demands a relatively large droop constant to compensate for the slow dynamics of the prime mover. A large droop constant in turn increases steady-state frequency deviation for a given load change, forcing system operators to maintain droop settings one order of magnitude higher than desirable for tight frequency control and creating heavy reliance on secondary regulation which needs time to act. Physical inertia is therefore considered essential because it ``buys time'' for secondary control \cite{Milano2018}.
	
	Power-electronic GFM converters face no such mechanical constraints: their time constants can be orders of magnitude faster, and the very real `virtual' damping they deliver is inversely proportional to their droop constant \cite{Arco2014}. It is shown that GFM systems invert a classical trade-off, fulfilling that smaller droop constants simultaneously provide stronger damping and stiffer frequency regulation, while virtual inertia becomes a freely tunable parameter rather than an intrinsic characteristic of the machine.
	
	This paper uses state-space modelling and modal analysis to demonstrate that the widely adopted practice of emulating legacy synchronous-machine behaviour (high virtual inertia, large droop) is a bad choice because it under-utilizes the potential of GFM converters. This is proven at a fundamental level in one and two-machine scenarios, which reveals an underlying nature of the interactions between GFM converters and conventional GC-SGs.
	
	Results show that replacing a GC-SG with a droop-based GFM converter of equivalent parameters already improves damping of the swing and turbine-governor modes. Further and counter-intuitive: deliberately selecting lower virtual inertia in the dramatically enhances electromechanical damping. These findings reveal that optimal GFM operation may lie in regions with low droop and low virtual inertia constants — yielding tightly bounded frequency variations, excellent damping of electromechanical modes, and a drastic reduction in the reliance on power system stabilizers (PSS) and secondary regulation.
	
	Current regulatory trends are leaning towards mandatory markedly inertial behaviour for GFM technology \cite{Lin2020,UNIFI2024,AEMO2023,ENTSOE2024}. The results of this study clearly point in the opposite direction and suggest that inertial behaviour must remain a degree of freedom in the optimization of each connection, with a continuous trend to be reduced as the electrical system evolves.
	
	Some researchers have already touched on the topic, reaching conclusions that align with ours, showing that the true meaning of inertia and its dynamic implications are non-intuitive and may be widely misunderstood. Reference \cite{Markovic2021} states that ``Contrary to popular belief, low inertia on its own does not have a major impact on the small-signal stability of power systems with high shares of PE-interfaced generation''. Work \cite{Pan2020} aligns with some of the claims of this paper: ``A small droop gain and a fast LPF with a high cutoff frequency (a low inertia) have been found critical for the active power loop'', but misses to identify the source of the confusion, originated in the structural differences between conventional synchronous machines and droop-based PE-GFM. Other authors \cite{Harnefors2019} have formulated upper, \textit{not lower} limits for the inertia constant in GFM devices to guarantee robust stability.
	
	Building upon these observations, this paper employs fundamental modal analysis to demonstrate why the trade-offs governing synchronous machines do not apply to GFM converters. It aims to clarify the fundamental structural differences between the two technologies and provide a theoretical basis for optimizing GFM parameters beyond the constraint of minimum inertial behaviour.
	
	The document is organized as follows. Section \ref{sec:assumptions} contains modeling assumptions. Section \ref{sec:microgrid_gcsg} derives the state-space model and characteristic equation for a single-machine microgrid with a single GC-SG. Section \ref{sec:microgrid_gfm} presents the equivalent droop-based GFM topology and highlights the inverted damping-droop relationship. Section \ref{sec:two_machine} extends the analysis to a two-machine system and performs modal analysis under different conditions. Section \ref{sec:rocof} examines the implications of the proposed alternatives on the Rate of Change of Frequency (RoCoF), discussing the trade-off between a larger instantaneous RoCoF and narrower overall frequency deviations. The analysis foundations and modeling principles closely align with \cite{Kundur1994}. Section \ref{sec:conclusions} concludes that current ``do-no-harm'' replacement policies (matching legacy $H$ and $R$ settings) may be suboptimal and advocates a paradigm shift in GFM control design and grid-code requirements.
	
	\section{Modeling Assumptions}
	\label{sec:assumptions}
	The analysis presented is based on the following assumptions, which are standard in studies of electromechanical dynamics and small-signal rotor angle/frequency stability:
	
	\textit{In general:}
	\begin{enumerate}
		\item Focus is placed exclusively on slow electromechanical dynamics (0.1--5\,Hz range).
		\item Models are linearized around steady-state operating points; small-disturbance linear simulations are used.
	\end{enumerate}
	
	\textit{For the Governor-Controlled Synchronous Generator (GC-SG):}
	\begin{enumerate}
		\setcounter{enumi}{2} 
		\item Fast electrical dynamics (stator transients, sub transient saliency) are disregarded.
		\item Voltage Control (AVR) is assumed ideal (i.e. machine terminal voltage magnitude is constant).
		\item The Power System Stabilizer (PSS) is not considered to purposely illustrate the swing mode.
		\item Loads are modelled as constant shunt impedances (primarily resistive in base cases).
		\item Turbine-governor systems are represented by a simple first-order lag with time constant \Tg; higher-order governor details are omitted as they do not alter the fundamental insights.
	\end{enumerate}
	
	\textit{For the Power-Electronics based Grid Former (PE-GFM):}
	\begin{enumerate}
		\setcounter{enumi}{7} 
		\item The modulation stage and the voltage/current loops are assumed to be ideal (i.e. the control law dictated by the droop law is applied instantly).
	\end{enumerate}
	\begin{enumerate}
		\setcounter{enumi}{8} 
		\item The PE-GFM is connected to a large-enough DC-battery.
	\end{enumerate}
	
	These assumptions are deliberate and widely adopted in the literature \cite{Kundur1994,Dorfler2013}, to enable analytical tractability while retaining the essential physics of inertia, damping, droop, and synchronizing torque interactions.
	
	\section{A Simplified Microgrid with a Governor-Controlled System}
	\label{sec:microgrid_gcsg}
	
\begin{figure*}[!t]
		\centering
		\begin{minipage}{0.48\textwidth}
			\centering
			\includegraphics[width=\linewidth]{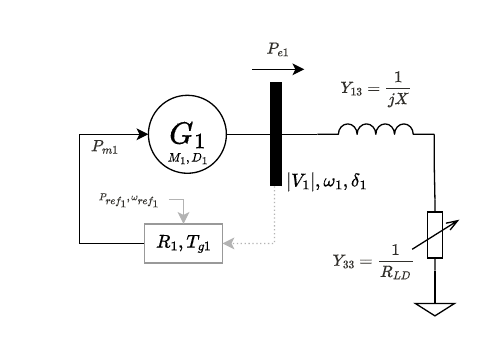}
			\caption{Configuration of a governor-controlled synchronous generator feeding a simplified load.}
			\label{fig:GCSG_model}
		\end{minipage}\hfill
		\begin{minipage}{0.48\textwidth}
			\centering
			\makebox[\linewidth][c]{\includegraphics[width=1.325\linewidth]{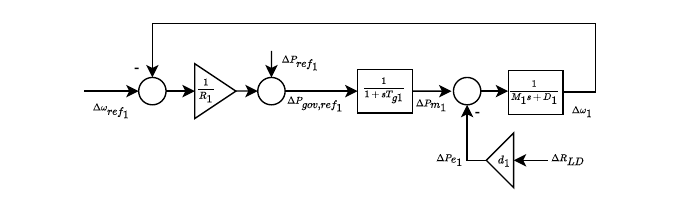}}
			\caption{Linearized small-signal block diagram of the governor-controlled synchronous generator.}
			\label{fig:GCSG_linear_model}
		\end{minipage}
	\end{figure*}

	Figure~\ref{fig:GCSG_model} represents a small-scale isolated power system with a governor-controlled genset feeding a load. The block diagram in Figure~\ref{fig:GCSG_linear_model}, reveals that at high frequencies the governor has zero gain, which makes the equivalent system respond to changes in electrical power with the inertia and the damping that are intrinsic to the nature of the generator ($M_1$, $D_1$). The large time constant of such a system is the reason why inertia makes electrical systems resilient. At very low frequencies the system response is dominated by the droop constant, which establishes the proportional relationship between speed and power variations that is fundamental to power synchronization.
	
	\begin{figure*}[!t]
		\centering
		\begin{minipage}{0.48\textwidth}
			\centering
			\includegraphics[width=\linewidth]{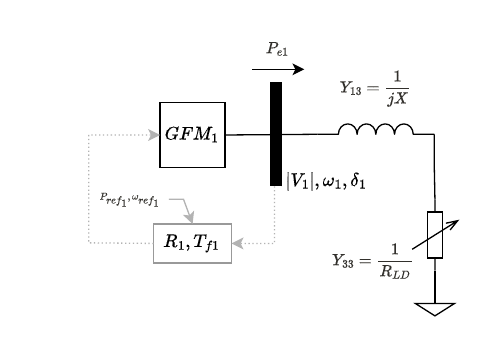}
			\caption{Droop-controlled grid-former feeding a load.}
			\label{fig:PEGFM_topology}
		\end{minipage}\hfill
		\begin{minipage}{0.48\textwidth}
			\centering
			\makebox[\linewidth][c]{\includegraphics[width=1.25\linewidth]{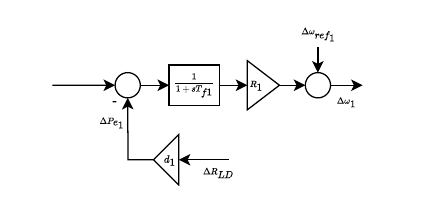}}
			\caption{Linearized small-signal block diagram of the PE-GFM converter.}
			\label{fig:PEGFM_model}
		\end{minipage}
	\end{figure*}
	
	The time derivative of the rotor angle ($\delta_i$, in rad) is the rotor speed deviation from the synchronous speed, $\Delta\omega_i$ in rad/s.
	\begin{equation}
		\frac{d\delta_1}{dt} = \Delta\omega_1 = \omega_1 - \omega_s
	\end{equation}
	
	The acceleration of the rotor is determined by the net torque (Newton's 2nd law), where following power system conventions, power is used instead of torque and angular momentum is used instead of momentum of inertia.
	\begin{equation}
		M_1 \frac{d\Delta\omega_1}{dt} = \Delta P_{m1} - \Delta P_{e1} - D_1 \Delta\omega_1
	\end{equation}
	$M_i$ is the angular momentum of the generator in MW$\cdot$s$^2$/rad ($M = J * \omega_s$), and $D_i$ is the generator damping coefficient in MW$\cdot$s/rad.
	
	The generator is equipped with a governor that varies the mechanical power setpoint, increasing it when the measured speed reduces, according to a droop constant $R_1$.
	\begin{equation}
		P_{gov\_ref,1} = P_{ref,1} + \frac{1}{R_1}(\omega_{ref,1} - \omega_1)
	\end{equation}
	
	The dynamics of the governor and the prime mover are represented with first order dynamics.
	\begin{equation}
		T_{g1} \frac{dP_{m1}}{dt} + P_{m1} = P_{gov\_ref,1}
	\end{equation}
	\begin{align}
		\frac{dP_{m1}}{dt} &= \frac{1}{T_{g1}}(-P_{m1} + P_{gov\_ref,1}) \nonumber \\
		&= \frac{1}{T_{g1}}\left(-P_{m1} + P_{ref,1} + \frac{1}{R_1}(\omega_{ref,1} - \omega_1)\right)
	\end{align}
	So,
	\begin{equation}
		T_{g1} \frac{d\Delta P_{m1}}{dt} = -\Delta P_{m1} + \Delta P_{ref1} - \frac{1}{R_1}(\Delta\omega_1 - \Delta\omega_{ref1})
	\end{equation}
	
	The system states are $x = [\Delta\omega_1, \Delta P_{m1}]^T$. The equation for electrical power needs to be linearized to obtain the complete SS representation. Notice, however, that it will not affect the dynamics matrix because it is not related to any of the states of the system.
	\begin{equation}
		P_{e1} = \frac{|V_1|^2}{R_{LD}^2 + X^2} R_{LD}
	\end{equation}
	\begin{equation}
		\Delta P_{e1} = \left.\frac{\partial P_{e1}}{\partial R_{LD}}\right|_{R_{LD,0}} \Delta R_{LD} = d_1 \Delta R_{LD}
	\end{equation}
	\begin{equation}
		d_1 = \frac{|V_1|^2}{(R_{LD,0}^2 + X^2)^2}(X^2 - R_{LD,0}^2)
	\end{equation}
	The system matrix is:
	\begin{equation}
		A = \begin{bmatrix}
			-D_1/M_1 & 1/M_1 \\
			-1/R_1 T_{g1} & -1/T_{g1}
		\end{bmatrix}
	\end{equation}
	
	\vspace{5pt}
	
	With characteristic equation:
	\begin{equation}
		\lambda^2 + \left(\frac{D_1}{M_1} + \frac{1}{T_{g1}}\right)\lambda + \left(\frac{D_1}{M_1 T_{g1}} + \frac{1}{M_1 R_1 T_{g1}}\right) = 0
	\end{equation}
	
	\vspace{5pt}
	
	Dropping the generator index and using per unit notation (Appendix~\ref{app:per_unit}):
	\begin{equation}
		\lambda^2 + \left(\frac{D_{pu}}{2H} + \frac{1}{\Tg}\right)\lambda + \left(\frac{D_{pu}}{2H\Tg} + \frac{1}{2HR_{pu}\Tg}\right) = 0
	\end{equation}
	
	\vspace{5pt}
	
	By analogy with $\lambda^2 + 2\zeta\omega_n\lambda + \omega_n^2 = 0$, the above-equation typically has a single, damped oscillatory mode known as the \textit{turbine-governor} mode. For conventional synchronous machines $2H \gg D_{pu}$ and $\frac{D_{pu}}{2H} \ll \frac{1}{\Tg}$, so the following approximations hold:
	
	\begin{itemize}
		\item Natural frequency:
		\begin{equation}
			f_n = \frac{1}{2\pi}\sqrt{\frac{1}{2H\Rpu\Tg}}
		\end{equation}
		
		\item Damping ratio:
		\begin{equation}
			\zeta = \sqrt{\frac{H\Rpu}{2\Tg}}
			\label{eq:damping ratio}
		\end{equation}
		
		\item The complex pole becomes critically damped when:
		\begin{equation}
			\Rpu > \frac{\Tg}{H}
		\end{equation}
	\end{itemize}
	
	\vspace{5pt}
	
	These equations reveal a crucial trade-off: to gain adequate damping, $\Rpu$ must be large enough to compensate for the sluggishness of the governor, which in turn makes the frequency more sensitive to load changes. That is, $\Rpu$ cannot be made arbitrarily large to gain damping because the operational band of frequency would easily be trespassed. Being the inertia constant a fixed parameter of the system, only increasing the governor dynamics (reducing $\Tg$) can enhance the damping of this mode. 
	
	Table~\ref{tab:turbine_gov_params} reveals the limits of governor technology, hardly ever faster than $T_g = 0.2$\,s. Poorly tuned governors may well be in the 2--10\,s range, and these are not uncommon in microgrids that do not require grid-code compliance. Combined with the fact that $H$ is given, this forces the choice of $R_{pu}$ (i.e. droop constant) to be above a minimum limit if adequate damping is sought for this mode. Conventional power systems are managed within narrow bands around their central frequency, which would require droop constants about one order of magnitude smaller of what the best-performing conventional governor technology can deliver, creating dependency on secondary regulation and the associated large inertia that buys time for it to operate. 
	
	\begin{table}[!t]
		\caption{Typical Turbine-Governor parameters and modal frequency, tuned for critical damping.}
		\label{tab:turbine_gov_params}
		\centering
		\begin{tabular}{lcccc}
			\toprule
			\textbf{Type} & \textbf{$T_g$ (s)} & \textbf{$H$ (s)} & \textbf{$R_{pu}$} & \textbf{$f_n$ (Hz)} \\
			\midrule
			Hydro & 0.2--0.5 & 3.0--9.0 & 0.022--0.167 & 0.23--0.56 \\
			Steam & 0.2--0.3 & 4.0--10.0 & 0.020--0.075 & 0.38--0.56 \\
			Gas/Genset & 0.1--0.3 & 5.0--9.0 & 0.011--0.060 & 0.38--1.13 \\
			Nuclear & 0.2--0.4 & 5.0--8.0 & 0.025--0.080 & 0.28--0.56 \\
			Coal-fired & 0.2--0.3 & 4.0--8.0 & 0.025--0.075 & 0.38--0.56 \\
			\bottomrule
		\end{tabular}
	\end{table}
	
	For the moment, recall that in a GC-SG a larger droop constant increases damping, but it also creates a stronger need for secondary regulation.
	
	\section{A Simplified Microgrid with a PE-GFM Converter}
	\label{sec:microgrid_gfm}
	
	The same exercise can be carried out with the core algorithm of a PE-GFM converter, Figures~\ref{fig:PEGFM_topology} and \ref{fig:PEGFM_model}. Comparing Figure~\ref{fig:GCSG_linear_model} and Figure~\ref{fig:PEGFM_model}, the reader may verify that when the generator damping is ignored (i.e. $D_1=0$), both structures reveal the same steady-state characteristic. Notably, the battery-fed grid-forming converter behaves like a rotating mass with both inertia and damping. It has been shown in literature \cite{Arco2014} that the droop law is equivalent to an inertial system. This can easily be shown by analogy:
	\begin{equation}
		\underbrace{\frac{1}{\frac{T_{f1}}{R_1}s + \frac{1}{R_1}}}_{\text{Inverter Control}} \equiv \underbrace{\frac{1}{(M_1 s + D_1)}}_{\text{Machine Physics}}
	\end{equation}
	This explains why this topology is often cited to deliver `virtual inertia' and `virtual damping':
	\begin{equation}
		M_{virtual} = \frac{T_{f1}}{R_1}, \quad D_{virtual} = \frac{1}{R_1}
	\end{equation}
	The droop constant $R_1$ is inversely proportional to damping. A small droop constant (stiff frequency control) results in high damping. This fact creates a significant misunderstanding on the role of droop controllers when applied to different technologies: \textit{For the PE-based GFM inverter, a smaller droop constant (stiffer response) leads to a larger damping coefficient, whereas for the governor-controlled turbine, this relationship works in the opposite direction.}
	
	The most relevant paradigm that this study challenges so far is that the choice of the droop constant does not anymore require to trade frequency stiffness for damping of the turbine-governor mode. By reducing the droop constant, any GFM system that successfully implements this control scheme will at the same time exhibit a larger inertia, reduce frequency variations in the network, reduce the need for secondary regulation, and increase damping.
	
	\section{A More Comprehensive System}
	\label{sec:two_machine}
	
	The introduction of a second generator is studied in this section. Note that the parameters of the governor-turbine topology, Figure~\ref{fig:GCSG_linear_model}, can be chosen to represent the PE-GFM if the governor droop constant, $R_1$ is chosen arbitrarily large. This effectively breaks the feedback loop, presenting to the network a pure inertial plus damping system.
	
	\begin{figure*}[!t]
		\centering
		\includegraphics[width=0.8\textwidth]{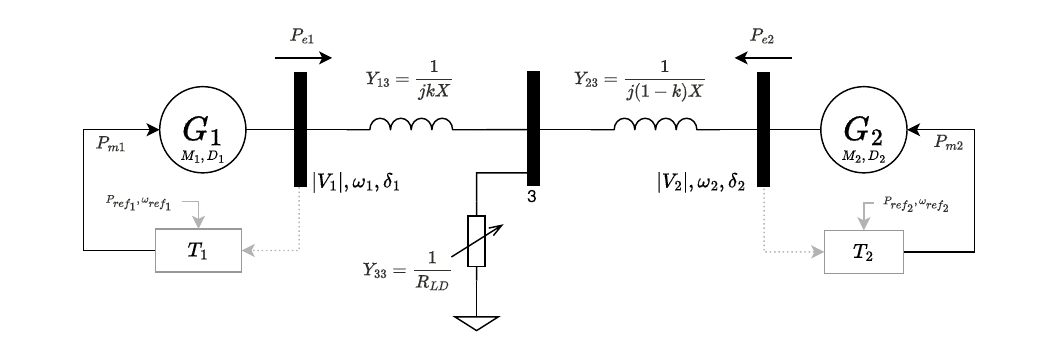}
		\caption{3-bus system with two droop-controlled generators and a load.}
		\label{fig:3bus_system}
	\end{figure*}
	
	\begin{figure*}[!t]
		\centering
		\includegraphics[width=0.8\textwidth]{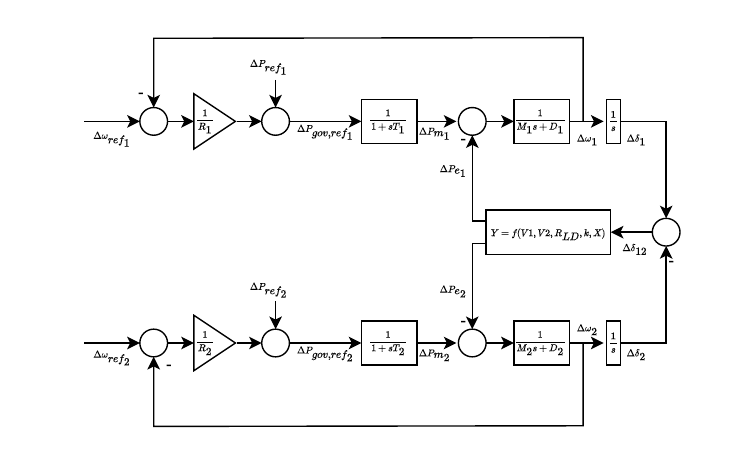}
		\caption{Linear diagram of the 3-bus system.}
		\label{fig:3bus_linear}
	\end{figure*}
	
	The development of this model is an extension of the previous, with the main difference that two generators are now coupled (Figure~\ref{fig:3bus_system}, Figure~\ref{fig:3bus_linear}). The expressions for the electrical power of each generator are (see Appendix~\ref{app:derivatives}):
\begin{align}
	P_{e1} &= |V_1|^2 G'_{11} + |V_1||V_2|(G'_{12} \cos\delta_{12} + B'_{12} \sin\delta_{12}) \label{eq:Pe1_main} \\
	P_{e2} &= |V_2|^2 G'_{22} + |V_1||V_2|(G'_{12} \cos\delta_{12} - B'_{12} \sin\delta_{12}) \label{eq:Pe2_main}
\end{align}
	The linearized system equations are:
	\begingroup
	\small 
	\begin{align}
		\frac{d\Delta\delta_{12}}{dt} &= \Delta\omega_1 - \Delta\omega_2 \\
		M_1 \frac{d\Delta\omega_1}{dt} &= \Delta P_{m1} - \Delta P_{e1} - D_1(\Delta\omega_1 - \omega_s) \\
		T_{g1} \frac{d\Delta P_{m1}}{dt} &= -\Delta P_{m1} + \Delta P_{ref1} - \frac{1}{R_1}(\Delta\omega_1 - \Delta\omega_{ref1}) \\
		M_2 \frac{d\Delta\omega_2}{dt} &= \Delta P_{m2} - \Delta P_{e2} - D_2(\Delta\omega_2 - \omega_s) \\
		T_{g2} \frac{d\Delta P_{m2}}{dt} &= -\Delta P_{m2} + \Delta P_{ref2} - \frac{1}{R_2}(\Delta\omega_2 - \Delta\omega_{ref2})
	\end{align}
	\endgroup
	The general state-space representation is given by:
	\begin{equation}
		\Delta\dot{x} = A\Delta x + B\Delta u, \quad \Delta y = C\Delta x + D\Delta u
	\end{equation}
	The state ($x$), input ($u$), and output ($y$) vectors are:
	\begingroup
	\begin{align}
		x &= [\Delta\delta_{12}, \Delta\omega_1, \Delta P_{m1}, \Delta\omega_2, \Delta P_{m2}]^T \nonumber \\
		u &= [\Delta P_{ref1}, \Delta P_{ref2}, \Delta\omega_{ref1}, \Delta\omega_{ref2}, \Delta R_{LD}]^T \\
		y &= [\Delta\omega_1, \Delta\omega_2, \Delta P_{e1}]^T \nonumber
	\end{align}
	\endgroup
	
	The linearized power coefficients represent the partial derivatives of the generator electrical power outputs with respect to the rotor angle difference and the load, evaluated at the equilibrium point. They show how changes in the relative angle between the two generators and how changes in the load affect the power of each generator. The calculation of these coefficients ($K_{lin1}, K_{lin2}, d_1, d_2$ ) is shown in Appendix~\ref{app:derivatives}.
	
	\begingroup
	\footnotesize
	\begin{align}
		\Delta P_{e1} &\approx \left.\frac{\partial P_{e1}}{\partial \delta_{12}}\right|_0 \Delta\delta_{12} + \left.\frac{\partial P_{e1}}{\partial R_{LD}}\right|_0 \Delta R_{LD} := K_{lin1} \Delta\delta_{12} + d_1 \Delta R_{LD} \\
		\Delta P_{e2} &\approx \left.\frac{\partial P_{e2}}{\partial \delta_{12}}\right|_0 \Delta\delta_{12} + \left.\frac{\partial P_{e2}}{\partial R_{LD}}\right|_0 \Delta R_{LD} := K_{lin2} \Delta\delta_{12} + d_2 \Delta R_{LD}
	\end{align}
	\endgroup
	The state-space matrices of the linearized model are:
	\begin{equation}
		\mbox{\small $\displaystyle
			\setlength{\arraycolsep}{2pt} 
			\renewcommand{\arraystretch}{2.00} 
			A = \left[\begin{matrix}
				0 & 1 & 0 & -1 & 0 \\
				-\frac{K_{lin1}}{M_1} & -\frac{D_1}{M_1} & \frac{1}{M_1} & 0 & 0 \\
				0 & -\frac{1}{R_1 T_{g1}} & -\frac{1}{T_{g1}} & 0 & 0 \\
				-\frac{K_{lin2}}{M_2} & 0 & 0 & -\frac{D_2}{M_2} & \frac{1}{M_2} \\
				0 & 0 & 0 & -\frac{1}{R_2 T_{g2}} & -\frac{1}{T_{g2}}
			\end{matrix}\right]
			$}
	\end{equation}	
	\begin{equation}
		\mbox{\small $\displaystyle
			\setlength{\arraycolsep}{2pt} 
			\renewcommand{\arraystretch}{2.00} 
			B = \left[\begin{matrix}
				0 & 0 & 0 & 0 & 0 \\
				0 & 0 & 0 & 0 & -\frac{d_1}{M_1} \\
				\frac{1}{T_{g1}} & 0 & \frac{1}{R_1 T_{g1}} & 0 & 0 \\
				0 & 0 & 0 & 0 & -\frac{d_2}{M_2} \\
				0 & \frac{1}{T_{g2}} & 0 & \frac{1}{R_2 T_{g2}} & 0
			\end{matrix}\right]
			$}
	\end{equation}
	
	\begin{equation}
		\mbox{ $\displaystyle
			\setlength{\arraycolsep}{2pt} 
			\renewcommand{\arraystretch}{1.2} 
			C = \left[\begin{matrix}
				0 & 1 & 0 & 0 & 0 \\
				0 & 0 & 0 & 1 & 0 \\
				K_{lin1} & 0 & 0 & 0 & 0
			\end{matrix}\right]
			$}
	\end{equation}
	
	\vspace{5pt}
	
	\begin{equation}
		\mbox{ $\displaystyle
			\setlength{\arraycolsep}{2pt} 
			\renewcommand{\arraystretch}{1.2} 
			D = \left[\begin{matrix}
				0 & 0 & 0 & 0 & 0 \\
				0 & 0 & 0 & 0 & 0 \\
				0 & 0 & 0 & 0 & d_1
			\end{matrix}\right]
			$}
	\end{equation}
	
	\vspace{5pt}
	
	\section{Modal Analysis of the Two-Generator System}
	The most fundamental way to analyse the system nature is to calculate its eigenvalues. These reveal oscillatory modes with associated frequencies and damping ratios. Table~\ref{tab:params_cases} contains parameters for three cases:
	
	\begin{table}[!t]
		\caption{Parameters for 3 cases. Case 1a: two GCSG. Cases 1b/1c: Generator \#1 emulates a GFM converter.}
		\centering
		\label{tab:params_cases}
		\begin{tabular}{lcccl}
			\toprule
			\textbf{Parameter} & \textbf{case 1a} & \textbf{case 1b} & \textbf{case 1c} & \textbf{Description} \\
			\midrule
			$P_{ref1},P_{ref2}$ & 0.50 & 0.50 & 0.50 & Dispatch (pu) \\
			\midrule
			$S_1,S_2$ & 0.5 & 0.5 & 0.5 & Rating (pu) \\
			$H_1,H_2$ & 4.0 & 4.0 & 4.0 & Inertia (s) \\
			$D_1$ & 0.01 & \textbf{20} & \textbf{100} & Damp. coeff. (pu) \\
			$D_2$ & 0.01 & 0.01 & 0.01 & Damp. coeff. (pu) \\
			$R_1$ & 0.05 & - & - & Droop const. (pu) \\
			$R_2$ & 0.05 & 0.05 & 0.05 & Droop const. (pu) \\
			$\tau_1$ & 0.25 & - & - & Gov. time const. (s) \\
			$\tau_2$ & 0.25 & 0.25 & 0.25 & Gov. time const. (s) \\
			\midrule
			SCR & 4 & 4 & 4 & Short-circuit ratio \\
			$k$ & 0.50 & 0.50 & 0.50 & $R_{LD}$ split point \\
			\bottomrule
		\end{tabular}
	\end{table}
	
	\vspace{5pt}	
	
	\begin{itemize}
		\item \textbf{Case 1a:} Weak connection supplied by two identical governor-controlled synchronous generators.
		\item \textbf{Case 1b:} GC-SG \#1 is replaced by a PE-GFM converter. The inertia (virtual) is maintained. The governor droop constant of the GC-SG it replaces is carried over to the GFM parameter $D_1 = 1/0.05 = 20$ pu.
		\item \textbf{Case 1c:} The GFM converter is pushed to deliver more damping (setting its droop constant to 0.01 pu).
	\end{itemize}
	
	\vspace{5pt}
	
	The operating point for all cases in Table~\ref{tab:params_cases} is $|V_3|=0.9659$ pu, $R_{ld}=0.9330$ pu, $\delta_{13}=\delta_{23}=15.0^\circ$.
	
	\begin{table}[!t]
		\caption{Modal analysis. Cases 1b and 1c contain a PE-GFM converter}
		\label{tab:modal_analysis}
		\centering
		\begin{tabular}{llllr}
			\toprule
			\textbf{Mode} & \textbf{Case} & \textbf{Eigenvalue} & \textbf{Freq (Hz)} & \textbf{$\zeta$} \\
			\midrule
			\multirow{3}{*}{\#1 Swing} & 1a & $-0.118 \pm 12.476j$ & 1.986 & 0.009 \\
			& 1b & $-0.645 \pm 12.233j$ & 1.947 & 0.053 \\
			& 1c & $-2.042 \pm 10.711j$ & 1.705 & 0.187 \\
			\midrule
			\multirow{3}{*}{\#2 Turb-Gov} & 1a & $-2.001 \pm 2.450j$ & 0.390 & 0.632 \\
			& 1b & $-2.605 \pm 1.727j$ & 0.275 & 0.833 \\
			& 1c (real) & $-7.461, -4.957$ & --- & --- \\
			\midrule
			\multirow{3}{*}{\#3 Gov} & 1a & $-3.766$ & --- & --- \\
			& 1b & --- & --- & --- \\
			& 1c & --- & --- & --- \\
			\bottomrule
		\end{tabular}
	\end{table}
	
	Table~\ref{tab:modal_analysis} shows that the system shown in Figure~\ref{fig:3bus_system} has three modes:
	
	\begin{itemize}
		\item \textbf{Mode 1} is the classical \textit{swing mode} (see Appendix~\ref{app:swing}), which is inherent to two rotating masses connected by a flexible shaft (the electrical line tie with reactance $X$). Its damping ratio, $\zeta$ is extremely small. This mode is often the focus of stability studies and cannot be significantly damped with any of the parameters that are part of the control systems which are in place in Case 1a. Under real-world conditions this mode would require damper-windings and a specifically-tuned PSS.
		
		\item \textbf{Mode 2} is the \textit{turbine-governor mode} (see Section~\ref{sec:microgrid_gcsg}), discussed already. This very low-frequency, better-damped mode is related to how mechanical power changes in response to changes in rotor speed. The ratio of the droop constant to the governor time constant, \eqref{eq:damping ratio}) is the main mechanism to increase the damping of this mode. In base Case 1a, the mode has sub-critical but acceptable damping ratio.
		
		\item \textbf{Mode 3} represents the \textit{internal governor control system}. It is real and it corresponds to approximately the inverse of the inertia-weighted average of the governor time constants. This approximation becomes less so as the network coupling is stronger (large SCR $\rightarrow$ eigenvalue at precisely $-1/T_{g1} = -4$). This mode disappears in Cases 1b and 1c.
	\end{itemize}
	
	\vspace{5pt}
	The physical damping coefficients which are inherent to each generator ($D_1,D_2$) are unable to deliver any significant contribution to the damping ratio of any of the modes. This has been verified well beyond any reasonable span of this parameter for Case 1a (i.e. 0.005 .. 0.05 pu).
	
	However, if one of the GC-SG is replaced by a PE-GFM device, as in Cases 1b and 1c, the damping that can be offered dramatically increases. Case 1b shows that, while keeping the same droop and inertia constants of the GCSM it replaces, the damping of the swing mode is increased five-fold. If the GFM is configured with reduced droop constant, as in Case 1c, the damping ratio of the swing mode improves considerably. 
	
	\textit{The substitution of a governor-controlled synchronous machine by a GFM converter with equivalent parameters significantly improves the damping ratio of the swing and turbine-governor modes.}
	
\begin{figure*}[!t]
		\centering
		\begin{minipage}{0.48\textwidth}
			\centering
			\includegraphics[width=\linewidth]{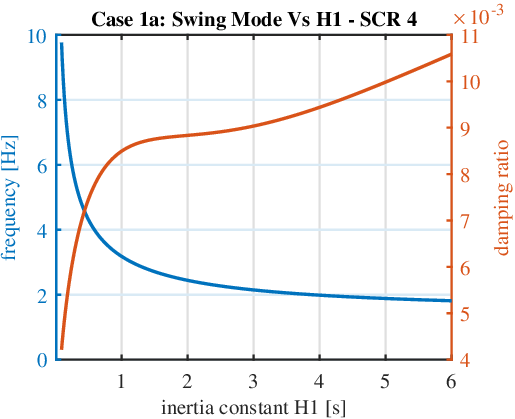}
		\end{minipage}\hfill
		\begin{minipage}{0.48\textwidth}
			\centering
			\includegraphics[width=\linewidth]{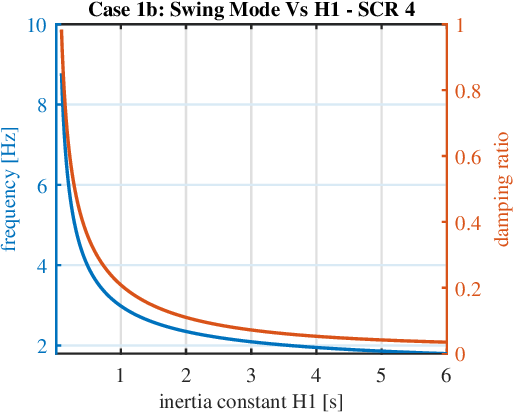}
		\end{minipage}
		\caption{Evolution of the swing mode when the inertia constant of generator \#1 varies freely. Left: conventional generation, Right: PE-GFM coupled to a conventional generator.}
		\label{fig:swing_mode_H}
	\end{figure*}

	For Case 1a, Figure~\ref{fig:swing_mode_H} Left, shows how varying $H_1$ results in the frequency of the swing mode following a predictable pattern, (Appendix~\ref{app:swing}), however the mode damping ratio remains stubbornly low.
	
	But if the same exercise is carried out for Case 1b, where one GC-SG has been replaced by a PE-GFM device, Figure~\ref{fig:swing_mode_H}, Right shows a large gain in mode damping as $H_1$ decreases, which is a practical possibility for a PE-GFM device.
	
	After the PE-GFM is set to instantiate the droop constant of the GC-SG it replaces, the natural choice would be to also replicate its inertia constant (i.e. 4s). But, as shown in Figure~\ref{fig:swing_mode_H}, Right, that choice would be a bad one: notice how \textbf{decreasing the inertia constant of the PE-GFM dramatically benefits the damping ratio of the swing mode}.
	
	\vspace{10pt}
	
	\section{Small-signal time-domain simulation}
	
	A new set of cases is prepared (Table~\ref{tab:time_domain_params}). These are made asymmetric by choosing $k \neq 0.5$: in a purely symmetrical system, differential modes, such as the swing mode, will not be excited by a symmetrical change in the load and thus they will not be observable. The governor time constant is purposely increased to worsen the damping ratio of the associated mode so that the low-frequency governor-turbine mode becomes more apparent.
	
	Base Case 2a contains two GC-SG while in Cases 2b and 2d machine \#1 has been replaced by a PE-GFM. In Case 2b, the GFM instantiates the legacy behaviour of the GCSM it emulates, delivering already noticeable improvements in mode damping. In Case 2d, the GFM converter is instructed to deliver very small inertia, showing remarkably enhanced dynamics.
	
	\begin{table}[!t]
		\caption{Parameters for time-domain simulation.}
		\centering
		\label{tab:time_domain_params}
		\begin{tabular}{lcccl}
			\toprule
			\textbf{Parameter} & \textbf{case 2a} & \textbf{case 2b} & \textbf{case 2d} & \textbf{Description} \\
			\midrule
			$P_{ref1},P_{ref2}$ & 0.50 & 0.50 & 0.50 & Dispatch (pu) \\
			\midrule
			$S_1,S_2$ & 0.5 & 0.5 & 0.5 & Rating (pu) \\
			$H_1$ & 4.0 & 4.0 & \textbf{0.02} & Inertia (s) \\
			$H_2$ & 4.0 & 4.0 & 4.0 & Inertia (s) \\
			$D_1$ & 0.01 & \textbf{20} & \textbf{20} & Damp. coeff. (pu) \\
			$D_2$ & 0.01 & 0.01 & 0.01 & Damp. coeff. (pu) \\
			$R_1$ & 0.05 & - & - & Droop const. (pu) \\
			$R_2$ & 0.05 & 0.05 & 0.05 & Droop const. (pu) \\
			$\tau_1$ & \textbf{1.0} & - & - & Gov. time const. (s) \\
			$\tau_2$ & \textbf{1.0} & \textbf{1.0} & \textbf{1.0} & Gov. time const. (s) \\
			\midrule
			SCR & 4 & 4 & 4 & Short-circuit ratio \\
			$k$ & \textbf{0.25} & \textbf{0.25} & \textbf{0.25} & $R_{LD}$ split point \\
			\bottomrule
		\end{tabular}
	\end{table}
	
	\begin{table}[!t]
		\caption{Modal analysis for cases 2a, 2b and 2d}
		\centering
		\label{tab:modal_analysis_case2}
		\begin{tabular}{lllll}
			\toprule
			\textbf{Mode} & \textbf{Case} & \textbf{Eigenvalue} & \textbf{Freq (Hz)} & \textbf{$\zeta$} \\
			\midrule
			\multirow{3}{*}{\#1 Swing} & 2a & $-0.126 \pm 10.209j$ & 1.625 & 0.001 \\
			& 2b & $-0.675 \pm 10.044j$ & 1.602 & 0.067 \\
			& 2d (real) & $-21.5, -476.3$ & --- & --- \\
			\midrule
			\multirow{3}{*}{\#2 Turb-Gov} & 2a & $-0.501 \pm 1.500j$ & 0.252 & 0.316 \\
			& 2b & $-1.075 \pm 1.164j$ & 0.252 & 0.679 \\
			& 2d & $-1.600 \pm 1.552j$ & 0.355 & 0.718 \\
			\midrule
			\multirow{3}{*}{\#3 Gov} & 2a & $-0.976$ & --- & --- \\
			& 2b & --- & --- & --- \\
			& 2d & --- & --- & --- \\
			\bottomrule
		\end{tabular}
	\end{table}
	
	\begin{figure*}[!t]
		\centering
		\begin{minipage}{0.32\textwidth}
			\centering
			\includegraphics[width=\linewidth]{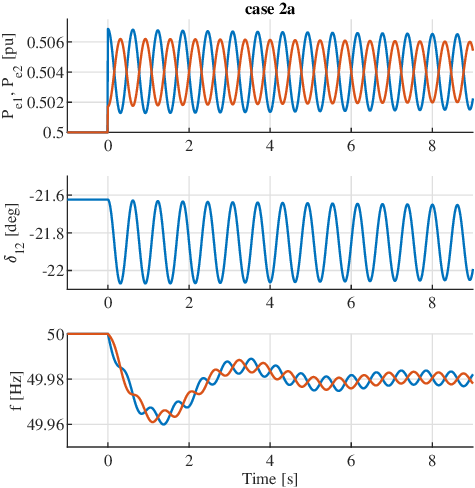}
		\end{minipage}
		\hfill
		\begin{minipage}{0.32\textwidth}
			\centering
			\includegraphics[width=\linewidth]{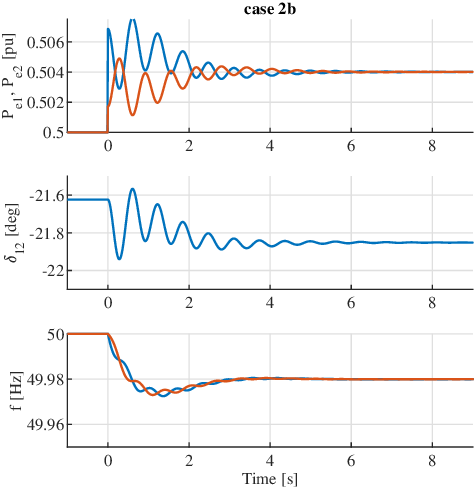}
		\end{minipage}
		\hfill
		\begin{minipage}{0.32\textwidth}
			\centering
			\includegraphics[width=\linewidth]{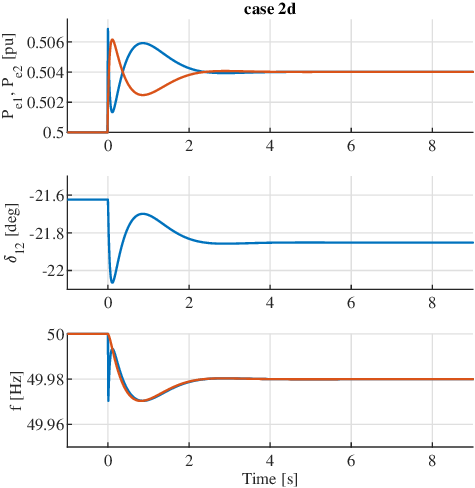}
		\end{minipage}
		\caption{Simulation results. Response to $\Delta R_{LD} =$ -1\%. Two coupled generators. (Left) Case 2a: GC-SG+GC-SG, (Center) Case 2b: PE-GFM (legacy) + GC-SG, (Right) Case 2d: PE-GFM (low H) + GC-SG.}
		\label{fig:simulation_results}
	\end{figure*}
	
	The simulations (Figure~\ref{fig:simulation_results}) illustrate the nature of the different modes. The swing-mode (of about 1.6\,Hz) is the visible ripple in Case 2a. It is a \textit{differential mode} because each system swings in nearly perfect opposite phase to its counterpart. The swing mode has nearly zero damping. In practice, damper windings and dedicated Power System Stabilizers (PSS) acting on the governor and AVR are used to damp this mode. This mode is still present when one synchronous machine is substituted by a PE-GFM converter that replicates its legacy parameters (Case 2b), but the dynamics are improved due to the additional damping that is achieved with the new control topology. Further, Case 2d shows the  dynamics once the PE-GFM converter is set to deliver very low inertia. Here, the swing mode has become faster and real, that is, irrelevant, as seen in Table~\ref{tab:modal_analysis_case2}. The reduced inertia makes the PE-GMF able to instantaneously adapt its phase to accomodate the load transient, as seen in the frequecy spike that only the PE-GMF experiences.
	
	The turbine-governor mode (Mode 2) is a \textit{common mode}, clearly visible in the frequency of Case 2a, due to its poor damping. Both systems oscillate together at 0.25\,Hz. This mode gets progressively improved in Cases 2b and 2d.
	
\begin{figure}[!t]
		\centering
		\includegraphics[width=0.8\linewidth]{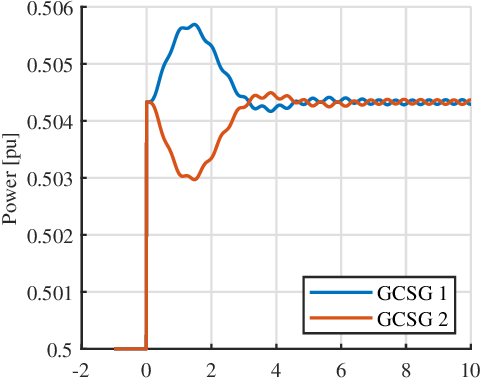}
		\caption{Illustration of the differential Governor mode when $G_1$ and $G_2$ have governors of different dynamics}
		\label{fig:gov_mode}
	\end{figure}

	Finally, the governor mode (Mode 3) is another \textit{differential mode} that arises from the interaction between the two governors. Only Case 2a presents this mode, but it is not immediately apparent under the current configuration: being a differential mode, there should be a marked difference each of the governor time constants GCSM for it to be noticeable. Its decay rate is the inverse of the weighted average the governors' time constants. For completion, Figure~\ref{fig:gov_mode} shows a symmetrical system ($k = 0.5$) where, $T_{g1} = 0.5$\,s and $T_{g2} = 1.5$\,s. They average to $1$\,s, so the mode is expected to settle (95\%) in about 3\,s.
	
	\section{On RoCoF}
	\label{sec:rocof} 
	
	The inverse relationship between the Rate of Change of Frequency (RoCoF) and the inertia constant of a power system is a primary driver to enforce minimum inertia requirements upon GFM converters. 
	
	High RoCoF is the primary indicator that frequency stability is compromised as it indicates that a large frequency drift is imminent (low nadir if generation is lost, high zenith if load is lost). Power systems incorporate frequency and RoCoF protections to disconnect generation or load as needed.
	
	The arguments presented here introduce the possibility of configuring PE-GFM converters to align somewhere in between two power system paradigms. Table~\ref{tab:low_inertia_comparison} highlights the structural differences between a conventional system and a moderately Low-Inertia, Low-Droop alternative.
	
	\begin{table}[!t]
		\renewcommand{\arraystretch}{1.1}
		\centering
		\footnotesize
		\caption{Study: RoCoF and Nadir of Conventional and Low~$H$, Low~$R$ Power Systems Subjected to a $25\%$ Generation Loss}
		\label{tab:low_inertia_comparison}
		\begin{tabularx}{\columnwidth}{@{} l >{\centering\arraybackslash}X >{\centering\arraybackslash}X @{}}
			\toprule
			& \textbf{Conventional} & \textbf{Low~$H$, Low~$R$} \\
			& \textbf{Power System} & \textbf{Power System} \\
			\midrule
			Equivalent $H$ [s]                     & 4                  & 0.5 \\
			\addlinespace[1pt]
			Natural $R$ [pu]        & 100                   & 0.02 \\
			\addlinespace[1pt]
			Primary reg. $R$ [pu] & 0.05               & - \\
			\addlinespace[1pt]
			Secondary regulation               & Enabled             & Optional? (Enabled) \\
			\addlinespace[1pt]
			RoCoF [Hz/s] - 50ms                & 1.56                & 4.6  \\
			\addlinespace[1pt]
			RoCoF [Hz/s] - 500ms               & 1.32                & 0.5  \\
			\addlinespace[1pt]
			$\Delta f$ Nadir [Hz]                         & -0.88               & -0.25 \\
			\addlinespace[1pt]
			\bottomrule
		\end{tabularx}
	\end{table}
	
\begin{figure}[!t]
		\centering
		\includegraphics[width=0.8\linewidth]{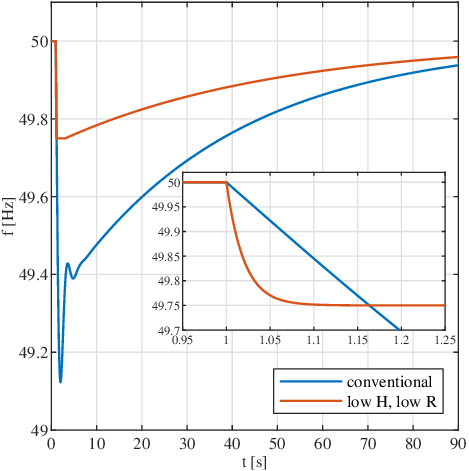}
		\caption{Frequency Nadir of a power system subjected to a generation loss of 25\%. The detail shows how a system with higher instantaneous RoCoF results in a significantly higher Nadir.}
		\label{fig:nadir}
	\end{figure}

	Figure~\ref{fig:nadir} is the time-domain simulation for the two configurations shown in Table~\ref{tab:low_inertia_comparison}. The blue trace represents a conventional power system with equivalent $H = 4$\,s, $R_{pu} = 0.05$\,pu, $T_g = 0.5$\,s (primary regulation), and natural damping $D_{pu} \approx 0.01$, utilizing slower secondary regulation to enforce isochronous operation. The red trace shows a PE-GFM dominated power system, configured to deliver high damping (equivalently, a low droop constant) and low inertia. 
	
	\textit{Despite presenting higher initial RoCoF, it is difficult to argue against the superior performance of the Low-Inertia, High Damping (Low R) alternative}. This illustrates that the concept of RoCoF becomes less meaningful as a system stress predictor when modern control alternatives are considered.
	
	The configuration of RoCoF protections is not trivial. Modern standards, such as \cite{IEEE1547_2018}, already make provision for RoCoF withstand capabilities larger than 3\,Hz/s in islanded systems and systems with a large proportion of IBR. These protections, while ensuring system reliability, must be robust against nuisance tripping, forcing their measuring windows to be larger than 0.1\,s. This measurement latency could facilitate the integration of the alternatives proposed here, as the large but extremely brief RoCoF spikes inherent to low-inertia GFM would effectively be filtered out by standard protection logic.
	
	\section{Conclusions}
	\label{sec:conclusions}
	Through state-space modelling and modal analysis, this paper has demonstrated that enforcing PE-GFM converters to emulate the dominant inertial characteristics of legacy synchronous generators may not be the best way forward. Droop-based PE-GFM control removes the historical trade-off that forces large droop constants for adequate turbine-governor mode damping. Further, it is shown that when a conventional GC-SG is coupled to a PE-GFM with reduced virtual inertia, it is straightforward to completely damp the swing mode and move its poles to higher frequency regions where it stops being a problem.
	
	This analysis reveals that the optimum usage of PE-GFM technology may lie far away from replicating the droop settings and inertial behaviour of legacy generators. Rather, it points to the opposite direction and suggests that the transformation of electrical systems should gradually target the reduction of the droop and inertia constants of the elements that form the grid. This will progressively eliminate inertial interactions and narrow the band in which the frequency can dwell and swing, and reduce the requirements on secondary regulation and power system stabilizers.
	
	The proposed low-inertia/low-droop/high-damping alternative inevitably results in larger instantaneous RoCoF values; however, these are transient and result in tightly bounded frequency deviations. This indicates that RoCoF may no longer be a valid metric to predict frequency Cenit/Nadir when modern PE-GFM alternatives are considered.
	
	As of November 2025, several jurisdictions explicitly or implicitly encourage/prescribe substantial virtual inertia provision (e.g., upcoming German rules, the new European Network Code, and ongoing AEMO, NERC/UNIFI discussions). The results presented here show that such requirements, if rigidly applied, may prevent the power system from realising the full benefits of power-electronic converters.
	
	\appendices
	\section{The Swing Mode}
	\label{app:swing}
	
	The swing mode that appears between two synchronous machines that are electrically coupled has a mechanical analogy with the textbook system comprised of two rotating-masses joined by a flexible shaft with friction. This section obtains this equivalence and provides more insight into the behaviour of the mode, which can easily be anticipated under certain conditions:
	
	\vspace{10pt}
	
	\begin{itemize}
		\item The voltage is healthy $\rightarrow |V_3| \approx |V_b| = |V_1| = |V_2|$
		\item The load is small compared to the network strength $\rightarrow R_{LD} \gg X$
		\item The generator damping is ignored $\rightarrow D_1 = D_2 \approx 0$
		\item The load is placed in the midpoint $\rightarrow k = 0.5, \delta_{12} \approx 0$
	\end{itemize}
	
	\vspace{5pt}
	
	The linearized coefficients become $K_{lin1} = -K_{lin2} \approx \frac{V^2}{X}$ and the swing equations become greatly simplified:
	
	\begin{equation}
		M_1 \frac{d\Delta\omega_1}{dt} \approx -\Delta P_{e1} \quad \text{and} \quad M_2 \frac{d\Delta\omega_2}{dt} \approx -\Delta P_{e2}
	\end{equation}
	From $\frac{d\Delta\delta_{12}}{dt} = \Delta\omega_1 - \Delta\omega_2$:
	
	\begin{equation}
		\begin{split}
			\frac{d^2\Delta\delta_{12}}{dt^2} &\approx -\left(\frac{V^2}{XM_1} + \frac{V^2}{XM_2}\right)\Delta\delta_{12} \\
			&= -\left(\frac{1}{M_1} + \frac{1}{M_2}\right)\frac{V^2}{X}\Delta\delta_{12}
		\end{split}
	\end{equation}
	
	Which corresponds to a second-order undamped system of natural frequency:
	
	\begin{equation}
		f_n = \frac{1}{2\pi}\sqrt{\left(\frac{1}{M_1} + \frac{1}{M_2}\right)\frac{V^2}{X}} = \frac{1}{2\pi}\sqrt{\frac{K_{eq}}{M_{eq}}}
	\end{equation}
	
	with $K_{eq} = \frac{V^2}{X}$ and $M_{eq} = M_1 || M_2$.
	
	\vspace{5pt}
	
\begin{figure}[!t]
		\centering
		\includegraphics[width=0.9\linewidth]{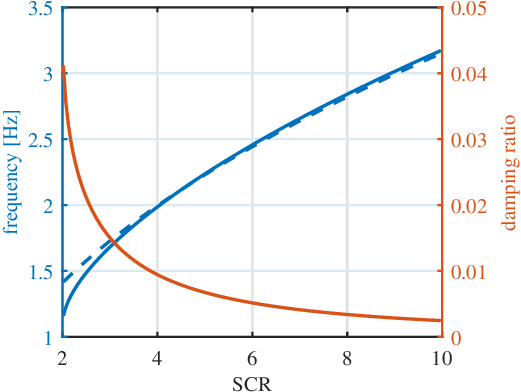}
		\caption{Swing mode dependency on SCR, symmetrical system with two generators with $H_1= H_2=1s$. (solid: full detail SS model, dashed: simplified prediction)}
		\label{fig:swing_scr}
	\end{figure}

	Figure~\ref{fig:swing_scr} shows the swing mode dependency on SCR. Crucially $H=1$\,s has been chosen, making the plot valid for any constant of inertia: the blue line can be scaled by $1/\sqrt{H}$ to obtain good approximations (i.e. if SCR = 8, and H = 4\,s, the mode will pulse at $5.686/\sqrt{4} = 2.84$\,Hz).
	
	The swing mode decreases its frequency as the network becomes weaker, making it difficult to damp as its frequency changes under different loading or network configurations. If the network is weak the amount of power which is pushed through it will drastically change the modal frequency. Therefore, conventional PSS tuning favours robustness over performance, and must be periodically retuned for network changes.
	
	Unless an effective voltage control is fitted in the midpoint, the voltage will rapidly collapse as more power is transferred to the load, as anticipated by \eqref{eq:|V3|}. Without this, below SCR = 2 the load flow does not even converge (i.e. there is not a physically valid solution to the problem). This should serve as a word of caution for connections with very low SCR.
	
	In the mechanical analogy (two rotating masses joined by a shaft), the stiffness of the `shaft' depends on the relationship between voltage squared and network reactance. For this reason, the frequency of the swing mode is proportional to the voltage and inversely proportional to the square root of the equivalent momentum of inertia of the network. This simplified calculation has been added with a dashed line to Figure~\ref{fig:swing_scr}.
	
	The damping coefficients of each generator ($D_1$ and $D_2$) do very little to dampen the mode. This is because this is a differential mode, ($\Delta\omega_1$ and $\Delta\omega_2$ oscillate against each other).
	
	\section{Calculation of Partial Derivatives}
	\label{app:derivatives}
	
	For the SS model $P_{e1}$ and $P_{e2}$ are required without dependencies on node 3 (i.e. $V_3, \delta_3$). To form the admittance matrix $Y$, diagonal elements ($Y_{ii}$) are obtained with the sum of all admittances connected to bus 'i'. Off-diagonal elements ($Y_{ij}$) are obtained with the negative of the admittance directly connecting bus 'i' and bus 'j'. Vector $I$ contains the currents injected by generators, and using conventional power flow notation, they are related to node voltages as follows:
	
	\begin{equation}
		I_{1,2,3} = Y_{3\times3}V_{1,2,3}
	\end{equation}
	
	\begin{equation}
		Y_{3\times3} = \begin{bmatrix}
			-\frac{j}{kX} & 0 & \frac{j}{kX} \\
			0 & -\frac{j}{(1-k)X} & \frac{j}{(1-k)X} \\
			\frac{j}{kX} & \frac{j}{(1-k)X} & \frac{1}{R_{LD}} - \frac{j}{k(1-k)X}
		\end{bmatrix}
	\end{equation}
	
	\subsection*{Kron reduction}
	
	The nodal admittance matrix $Y$ is $3\times3$, can be reduced to a $2\times2$ matrix using the Kron reduction method. The general formula for Kron reduction to eliminate node 'k' is: $Y'_{ij} = Y_{ij} - (Y_{ik} \cdot Y_{kj}) / Y_{kk}$. The node to be eliminated is $k=3$. The resulting $2\times2$ $Y$ matrix is: $Y_{red} = [Y'_{11}\ Y'_{12} ; Y'_{21}\ Y'_{22}]$.
	
	\vspace{5pt}
	
	\begin{align}
		Y'_{11} &= Y_{11} - (Y_{13} \cdot Y_{31}) / Y_{33} \nonumber \\
		Y'_{12} &= Y_{12} - (Y_{13} \cdot Y_{32}) / Y_{33} \nonumber \\
		Y'_{21} &= Y'_{12} \nonumber \\
		Y'_{22} &= Y_{22} - (Y_{23} \cdot Y_{32}) / Y_{33}
	\end{align}
	
	\begin{equation}
		Y_{red} = \begin{bmatrix}
			Y'_{11} & Y'_{12} \\
			Y'_{21} & Y'_{22}
		\end{bmatrix} = \begin{bmatrix}
			G'_{11} + jB'_{11} & G'_{12} + jB'_{12} \\
			G'_{21} + jB'_{21} & G'_{22} + jB'_{22}
		\end{bmatrix}
	\end{equation}
	
	The system equations are simplified. The system is reduced to a pi-equivalent.
	\begin{equation}
		I_{1,2} = Y_{2\times2}V_{1,2}
	\end{equation}
	
	\begin{figure}[!t]
		\centering
		\makebox[\linewidth][c]{\includegraphics[width=1.09\linewidth]{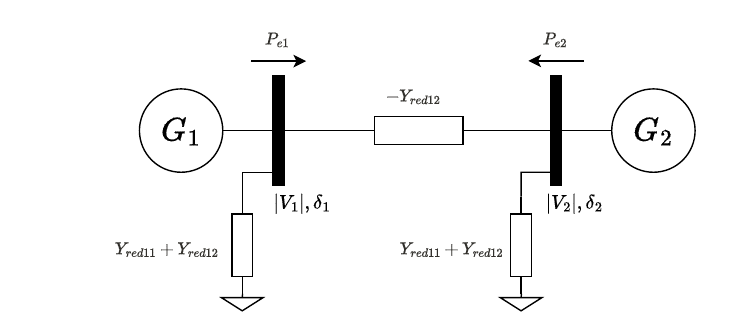}}
		\caption{Reduced pi-equivalent of the 3-bus system}
		\label{fig:reduced_pi}
	\end{figure}
	
Calculating the apparent power in $G_1$ and $G_2$ and taking the real part yields the expressions for $P_{e1}$ and $P_{e2}$ previously presented in \eqref{eq:Pe1_main} and \eqref{eq:Pe2_main}.
	
	With:
	\begin{equation}
		\delta_{12} = \delta_1 - \delta_2
	\end{equation}
	\begin{equation}
		\frac{d\delta_{12}}{dt} = \frac{d\delta_1}{dt} - \frac{d\delta_2}{dt} = \Delta\omega_1 - \Delta\omega_2
	\end{equation}
	
	The elements of $Y_{red} = [Y'_{11}\ Y'_{12} ; Y'_{21}\ Y'_{22}]$ are derived as follows. 
	
	\begin{align}
		\mbox{\small $\displaystyle Y'_{11}$} &= \mbox{\small $\displaystyle Y_{11}-\frac{Y_{13}Y_{31}}{Y_{33}} =  -\frac{j}{kX}-\frac{\frac{j}{kX}\frac{j}{kX}}{\frac{1}{R_{LD}}-\frac{j}{k(1-k)X}}$} \nonumber\\
		&= \mbox{\small $\displaystyle -\frac{j}{kX}+\frac{\frac{1}{k^2X^2}}{\frac{1}{R_{LD}}-\frac{j}{k(1-k)X}}$} \nonumber \\
		\mbox{\small $\displaystyle Y'_{12}$} &= \mbox{\small $\displaystyle Y_{12}-\frac{Y_{13}Y_{32}}{Y_{33}} = 0-\frac{\frac{j}{kX}\frac{j}{(1-k)X}}{\frac{1}{R_{LD}}-\frac{j}{k(1-k)X}}$} \nonumber\\
		&= \mbox{\small $\displaystyle \frac{\frac{1}{k(1-k)X^2}}{\frac{1}{R_{LD}}-\frac{j}{k(1-k)X}}$} \nonumber \\
		\mbox{\small $\displaystyle Y'_{22}$} &= \mbox{\small $\displaystyle Y_{22}-\frac{Y_{23}Y_{32}}{Y_{33}} = -\frac{j}{(1-k)X}-\frac{\frac{j}{(1-k)X}\frac{j}{(1-k)X}}{\frac{1}{R_{LD}}-\frac{j}{k(1-k)X}}$} \nonumber\\
		&= \mbox{\small $\displaystyle -\frac{j}{(1-k)X}+\frac{\frac{1}{(1-k)^2X^2}}{\frac{1}{R_{LD}}-\frac{j}{k(1-k)X}}$}
	\end{align}
	
	We define the term $\mathcal{D}$ to avoid confusion with damping coefficient $D$:
	
	\begin{equation}
		\mathcal{D} := \left(\frac{X}{R_{LD}}k(k-1)\right)^2 + 1
	\end{equation}
	
	The real and imaginary parts of each element in the admittance matrix are calculated:
	
	\begin{align}
		G'_{12} &= \frac{k(1-k)}{R_{LD}\mathcal{D}} \nonumber \\
		B'_{12} &= \frac{1}{X\mathcal{D}} \nonumber \nonumber \\
		G'_{11} &= \frac{(1-k)^2}{R_{LD}\mathcal{D}} \nonumber \\
		B'_{11} &= \frac{1 - k - \mathcal{D}}{kX\mathcal{D}} \nonumber \\
		G'_{22} &= \frac{k^2}{R_{LD}\mathcal{D}} \nonumber \\
		B'_{22} &= \frac{k - \mathcal{D}}{(1-k)X\mathcal{D}}
	\end{align}
	The equations for electrical power are rewritten to:
	
	\begin{equation}
		\begin{split}
			P_{e1} &= \frac{|V_1|}{\mathcal{D}}\bigg[ |V_1|\frac{(1-k)^2}{R_{LD}} \\
			&\quad + |V_2|\left(\frac{k(1-k)}{R_{LD}}\cos(\delta_{12}) + \frac{1}{X}\sin(\delta_{12})\right) \bigg]
		\end{split}
	\end{equation}
	
	\begin{equation}
		\begin{split}
			P_{e2} &= \frac{|V_2|}{\mathcal{D}}\bigg[ |V_2|\frac{k^2}{R_{LD}} \\
			&\quad + |V_1|\left(\frac{k(1-k)}{R_{LD}}\cos(\delta_{12}) - \frac{1}{X}\sin(\delta_{12})\right) \bigg]
		\end{split}
	\end{equation}
	
	The sine term is the classical synchronizing torque, acting as a restoring `spring' that pulls the relative rotor angle $\delta_{12}$ towards equilibrium. The cosine term, arising from the resistive load, modifies this behaviour in an operating-point-dependent manner: it reduces the synchronizing stiffness when one machine leads and increases it when it lags. This simplified network model has no inherent damping, but this is consistent with the study focus on synchronizing mechanisms. The governor, which adjusts mechanical power in response to speed deviations, is the primary source of damping for slow electromechanical modes of oscillation, as shown in \eqref{eq:damping ratio}
	
	\vspace{5pt}
	
	From $P_{e1} + P_{e2} = P_{LD} = \frac{|V_3|^2}{R_{LD}}$, the voltage at the load can be explicitly obtained:
	
	\begin{equation}
		\mbox{\small $\displaystyle
			|V_3| = \sqrt{\frac{|V_1|^2(1-k)^2 + |V_2|^2 k^2 + 2|V_1||V_2|k(1-k)\cos(\delta_{12})}{\mathcal{D}}}
			$}
		\label{eq:|V3|}
	\end{equation}
	
	\vspace{5pt}
	
	The numerator is the magnitude of the phasor sum of the two source voltages, weighted by their electrical distance to the load. The term $\sqrt{\mathcal{D}}$ represents the voltage-divider effect of the load. As the load gets heavier ($R_{LD}$ gets smaller), $\mathcal{D}$ increases, and the voltage $|V_3|$ "sags" or is pulled down. Thus, $\mathcal{D}$ is a measure of the voltage stiffness of this system. Large values of $R_{LD}$ ($\mathcal{D} \to 1$) indicate that the voltage profile will be flatter.
	
	\textbf{Partial derivatives with respect to $\delta_{12}$:} The elements of the reduced admittance matrix do not depend on $\delta_{12}$ and are shown with explicit dependencies on the admittance matrix elements.
	
	\begin{equation}
		\begin{split}
			K_{lin1} &:= \left.\frac{\partial P_{e1}}{\partial\delta_{12}}\right|_{\delta_{12,0}, P_{L,0}} \\
			&= |V_1||V_2|(-G'_{12,0}\sin\delta_{12,0} + B'_{12,0}\cos\delta_{12,0})
		\end{split}
	\end{equation}
	
	\begin{equation}
		\begin{split}
			K_{lin2} &:= \left.\frac{\partial P_{e2}}{\partial\delta_{12}}\right|_{\delta_{12,0}, P_{L,0}} \\
			&= |V_1||V_2|(-G'_{12,0}\sin\delta_{12,0} - B'_{12,0}\cos\delta_{12,0})
		\end{split}
	\end{equation}
	
	\vspace{5pt}
	
	After some manipulation, the \textbf{partial derivatives with respect to $R_{LD}$} are obtained:
	
	\begin{equation}
		\begin{split}
			d_1 &:= \left.\frac{\partial P_{e1}}{\partial R_{LD}}\right|_{R_{LD,0}, \delta_{12,0}} \\
			&= \frac{1}{\mathcal{D}_0 \cdot R_{LD,0}}\left[ P_{e1,0}(\mathcal{D}_0 - 2) + \frac{|V_1||V_2|}{X}\sin(\delta_{12,0}) \right]
		\end{split}
	\end{equation}
	
	\begin{equation}
		\begin{split}
			d_2 &:= \left.\frac{\partial P_{e2}}{\partial R_{LD}}\right|_{R_{LD,0}, \delta_{12,0}} \\
			&= \frac{1}{\mathcal{D}_0 \cdot R_{LD,0}}\left[ P_{e2,0}(\mathcal{D}_0 - 2) - \frac{|V_1||V_2|}{X}\sin(\delta_{12,0}) \right]
		\end{split}
	\end{equation}
	
	\section{Per Unit Notation}
	\label{app:per_unit}
	
	The mathematical modelling and state-space analysis are performed using physical SI units to ensure rigorous adherence to physical laws. However, parameters are defined in per-unit (pu) to align with standard industry practices.
	
	The base angular frequency is $\omega_b = 2\pi f_0$ [rad/s]. The base voltage, $V_{base}$ [V] is defined as the line-to-line RMS voltage, and the power base, $S_{base}$ [W] is chosen to be the sum of the ratings of the two generators.
	
	From those, the base impedance is:
	\begin{equation}
		Z_b = \frac{V_{base}^2}{S_{base}} \quad [\Omega]
	\end{equation}
	
	Each generator $i$ has a specific apparent power rating, $S_{n,i}$, defined as a fraction of the system base: $S_{n,i} = S_{i,pu} \cdot S_{base}$ [VA]. The dynamic coefficients for angular momentum ($M$), damping coefficient ($D$), and governor droop constant ($R$) are converted from their per-unit values (normalized to the machine rating $S_{n,i}$) into SI units for the simulation:
	
	\begin{equation}
		M_i = \frac{2H_i S_{n,i}}{\omega_b} \quad [W \cdot s^2/rad]
	\end{equation}
	
	\begin{equation}
		D_i = \frac{D_{i,pu} S_{n,i}}{\omega_b} \quad [W \cdot s/rad]
	\end{equation}
	
	\begin{equation}
		R_i = R_{i,pu}\frac{\omega_b}{S_{n,i}} \quad [(rad/s)/W]
	\end{equation}
	
	The per-unit notation allows using the SCR index to describe the strength of a system in a condensed way. Ours consists of two generators in parallel, supplying a load. The system base is chosen so that $P_{LD\_pu} = 1$ when each generator is supplying its rated power. For the calculation of the SCR the parallel (not the sum) of the two segments' impedances must be used.
	
	\begin{equation}
		SCR = \frac{1}{X_{pu\_parallel}} = \frac{1}{X_{pu}k(1-k)}
	\end{equation}
	

\end{document}